\documentclass[12pt]{iopart}
\usepackage{epsf}
\begin{document}
\title[Damping of vortex waves in a superfluid]
{Damping of vortex waves in a superfluid}
\author{H M Cataldo}
\address{Departamento de F\'{\i}sica, Facultad de Ciencias Exactas y
Naturales, Universidad de Buenos Aires, RA-1428 Buenos Aires, Argentina}
\ead{cataldo@df.uba.ar}
\begin{abstract}
The damping of vortex cyclotron modes is investigated within a generalized quantum theory
of vortex waves. Similarly to the case of Kelvin modes, the friction coefficient turns out
to be essentially unchanged under such oscillations, but it is shown to be affected by
appreciable memory corrections. On the other hand, the
nonequilibrium energetics of the vortex, which is
investigated
within the framework of linear response theory, shows that its memory corrections are negligible.
The vortex response is found to be
of the Debye type, with a relaxation frequency whose dependence on temperature and impurity
concentration reflects the complexity of the heat bath and its interaction with the vortex.
\end{abstract}
\submitto{\JPA}
\pacs{05.30.-d, 05.40.Jc, 67.40.-w, 67.60.-g}
\maketitle

\section{Introduction}
The simplest vortex dynamics in a superfluid corresponds to the two-dimensional motion of a 
rectilinear vortex filament \cite{don}. In fact, in an infinite superfluid 
a vortex ``charged" with one quantum of counterclockwise circulation will
move like an electron in a uniform magnetic field, ie performing a circular cyclotron
motion ruled by:
\begin{equation}
m_v{\bf \ddot{r}}=\rho_s h \hat{z}\times {\bf \dot{r}}.\label{1}
\end{equation}
Here $m_v$ denotes the vortex effective mass per unit length, $\rho_s$ the number density of the
background superfluid at rest, $h$ the Planck's constant and ${\bf r}=(x,y)$ the two-dimensional
coordinate of the vortex core. We note that the Magnus force in the right-hand side of 
\eref{1} is formally equivalent
 to the Lorentz force on a negative point charge in a uniform magnetic field
parallel to the $z$ axis. Actually, this electromagnetic analogy is only a part of a whole 
mapping by which a 2-D homogeneous superfluid can be mapped onto a (2+1)-D electrodynamic 
system, with vortices and phonons playing the role of charges and photons, respectively 
\cite{arovas}. For instance, any accelerated motion of a vortex would result in the radiation 
of sound waves in the superfluid, ie the emission of phonons, 
a process which is entirely analogous 
to the photon radiation mechanism stemming from an accelerated charge in electrodynamics.
In practice, however, this simple picture only would apply to a superfluid formed by $^4$He
atoms, viz the boson isotope of helium, at low temperatures ($T<$0.4 K). Ordinary helium,
on the other hand, contains a small amount of impurity, fermion $^3$He atoms, 
which however produces
a viscous drag force on a moving vortex at the lowest temperatures. In fact, below 0.4 K
the scattering of thermal phonons by the vortex has negligible effects compared to the drag
force due to $^3$He scattering \cite{reif}. In addition, at higher temperatures ($T>$0.5
K), most of the elementary excitations of the superfluid $^4$He that collide with the vortex 
are {\it rotons}, ie quasiparticles having their momentum around the minimum of the dispersion
curve. In fact, (see figure 1) only the elementary excitations with momentum below 0.5 \AA$^{-1}$
can be regarded as phonons arising from a linear dispersion relation $\omega=c_s k$, 
whereas the rest of the
dispersion curve do not yield thermal elementary excitations, except for a small interval 
around the minimum ($\sim$ 1.9 \AA$^{-1}$).
\begin{figure}
\begin{center}
\epsfbox{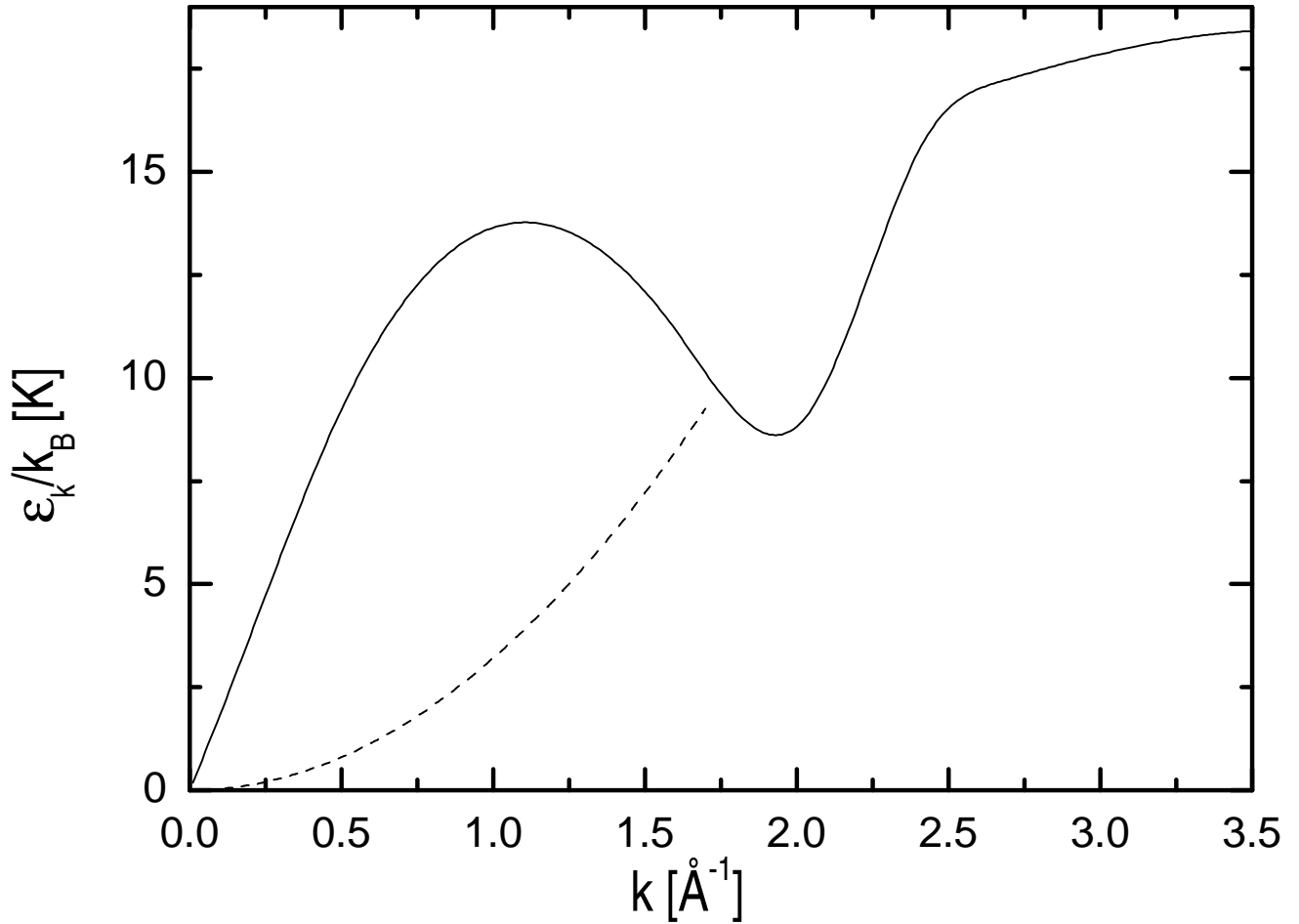}
\end{center}
\caption{\label{label}Dispersion curve for elementary excitations in superfluid $^4$He (full
curve) and energy spectrum of solvated $^3$He atoms in ordinary helium (broken curve).}
\end{figure}
 Quasiparticles with momentum at the right (left)
of this minimum have their group velocity parallel (antiparallel) to their momentum and are 
called $R^+$ ($R^-$) rotons. Above 0.5 K the source of the
 drag force on a vortex in ordinary helium is
 roton scattering, being the effect of $^3$He collisions practically negligible.
Whatever the source, however,
 such a drag force can be written as two additional terms in the right-hand
side of equation \eref{1}, viz
\begin{equation}
m_v{\bf \ddot{r}}=(\rho_s h-D') \hat{z}\times {\bf \dot{r}}-D\,{\bf \dot{r}},\label{p2}
\end{equation}
where $D'$ and $D$ denote {\it transversal} and {\it longitudinal} friction coefficients, 
respectively \cite{don,bar}.
 It is interesting to compare in figure 1
 the dispersion curve for elementary excitations in superfluid 
$^4$He, with the energy spectrum of $^3$He atoms in ordinary
helium. Such solvated atoms behave like heavier free particles ($\varepsilon_k=
\hbar^2k^2/2m^*$) with
an effective mass $m^*$ which exceeds two times the mass of a bare $^3$He atom. As a final 
remark about figure 1 we note that the terminations of
 both curves are due to unstabilities
caused by roton creation processes. That is, any elementary excitation exceeding two times
the roton energy should be unstable against decay into two rotons, whereas $^3$He atoms 
exceeding the roton energy should decay into a low energy atom plus a roton.

Here it is important also to take into account another consequence of the above scattering 
processes, apart from the friction itself, that is the thermal excitation of vortex waves
\cite{don,flu}. These are helical waves in which each vortex line element executes a circular
motion about the undisturbed line ($z$ axis). The radius of such a circle is assumed to be much
smaller than the wavelength, so the above elements will keep almost parallel to the $z$ axis,
fulfilling an equation of motion like \eref{1}. In fact, such an equation has to be generalized
to the situation where there exists an external superfluid flow of velocity ${\bf v}_s$:
\begin{equation}
m_v{\bf \ddot{r}}=\rho_s h \hat{z}\times ({\bf \dot{r}}-{\bf v}_s).\label{p3}
\end{equation}
Such an 'external' superflow corresponds in our case to the local self-induced velocity generated
by the vortex line curvature \cite{don,hama}, ${\bf v}_s=-v_i\,{\bf\hat{\theta}}$, which
points in a direction opposite to the one of the superfluid 
velocity field generated by the undisturbed vortex line. This self-induced velocity, being 
proportional to the line displacement $|{\bf r}|$ from the $z$ axis, can be written as $v_i=
\omega_-|{\bf r}|$, where
\begin{equation}
\omega_-(k)\simeq-\frac{\hbar k^2}{2m_4}[\ln(|k|a)+0.116]\label{p4}
\end{equation}
corresponds to the well-known dispersion relation for Kelvin waves of long wavelength
$\lambda$, $|k|a\ll1$, being $|k|=2\pi/\lambda$, $a\sim$ 1 \AA $\,=$ vortex core parameter and
$m_4$ = mass of a $^4$He atom. Notice that the wave vector $k$, which points along the $z$ axis,
can take positive or negative values depending on the two ways of generating the vortex helix.
Then, equation \eref{p3} can be rewritten
\begin{equation}
{\bf \ddot{r}}=\Omega\, \hat{z}\times {\bf \dot{r}}-\Omega\,\omega_-{\bf r},\label{p5}
\end{equation}
where
\begin{equation}
\Omega=\rho_sh/m_v\label{2}
\end{equation}
corresponds to the cyclotron frequency arising from the equation of motion \eref{1}. Actually,
it is easy to check that the equation
 \eref{p5} allows both possible directions for circular motion, since
a replacement $|{\bf r}|=$ const, ${\bf \dot{r}}=\omega\,|{\bf r}|\,{\bf\hat{\theta}}$ 
in \eref{p5} leads to
a quadratic equation in the angular frequency $\omega$ with solutions:
\begin{equation}
\label{p7}
\omega^{(\pm)}=\frac{\Omega}{2}\left[1\pm\sqrt{1+\frac{4\omega_-}{\Omega}}\right].
\end{equation}
That is, in the limit $\omega_-/\Omega\ll1$ we have either the counterclockwise cyclotron motion
of frequency $\omega^{(+)}\simeq\Omega$, or the usual clockwise polarization of Kelvin waves
$\omega^{(-)}\simeq-\omega_-$ (cf \eref{p4}), for vortices of counterclockwise circulation and
negligible mass ($\Omega\rightarrow\infty$). Actually, there is no
experimental data about vortex trajectories, so the value of the vortex mass $m_v$ and hence
of $\Omega$, 
can only be extracted from theoretical considerations. If $m_v$ is calculated from a classical
hydrodynamical model, $\Omega$ should be about 3 ps$^{-1}$ \cite{don}, whereas more recent
theories, for which $m_v$ should be logarithmically divergent with the system size, lead to
$\Omega$'s of order $0.1-0.01$ ps$^{-1}$, for typical experimental conditions 
\cite{duan,tang}. All these figures are consistent with the approximation $\Omega\gg\omega_-$
for long wavelengths $|k|a\ll1$, where the dependence on $k$ of the frequency $\omega^{(+)}$
can be neglected \cite{don}.

Finally, the vortex equation of motion is obtained by adding both effects, friction \eref{p2}
and oscillations \eref{p5} together:
\begin{equation}
{\bf \ddot{r}}=\Omega\left[\left(1-\frac{D'}{\rho_sh}\right)
 \hat{z}\times {\bf \dot{r}}-\frac{D}{\rho_sh}{\bf \dot{r}}-\omega_-{\bf r}\right].\label{p8}
\end{equation}
Note that in addition to the time dependence of ${\bf r}$, one should take into account a
parametric dependence ${\bf r}(z)$ following the helix curvature.

To this point we have regarded the vortex coordinates as classical time-dependent variables,
but it is important to observe in this respect that the above
theoretical estimates of the cyclotron frequency yield
values of $\hbar\Omega/k_BT$ greater than $\sim$0.1 for $T<1$ K. This seems to indicate that
a classical treatment cannot be wholly satisfactory. Moreover, even in the case of purely
low-frequency Kelvin waves, the need of a quantum mechanical analysis was early pointed out
by Fetter \cite{fet1}. Such a theory was in fact used to study phonon scattering by a vortex
\cite{fet2}, and it is our purpose to present in this article a more general treatment at 
which the vortex mass, and hence the cyclotron frequency, 
are included in the theory and assigned
finite values. Our starting point will be a vortex Hamiltonian from which the equation of 
motion \eref{p5} derives. Then, after quantization of the vortex variables, we 
will show that such
a Hamiltonian consists of independent harmonic oscillator modes of frequencies $\Omega$ and
$\omega_-$, which interact with the heat bath represented by the ordinary helium at
a finite temperature.
Such an interaction is modelled through a generic momentum-conserving scattering 
Hamiltonian, which is used to study the dissipative dynamics of the cyclotron modes. Thus, 
we shall show that the friction turns out to be essentially unaffected by such oscillations, 
allowing us to extend our previous conclusions on the memory effects on straight vortex lines
\cite{jltp}. A similar behaviour was long ago reported by Fetter \cite{fet2} and Sonin \cite{son}
for the low-frequency Kelvin modes, showing that the dissipation of such modes remains 
essentially equal to that of strictly rectilinear vortices.

Our main objective in the present paper will be to analyze, 
within the framework of linear response theory \cite{datta},
 the non-equilibrium dynamics and the
equilibrium quantum fluctuations of the energy of the cyclotron modes.
 There exists an extensive literature on
 quantal Brownian motion of harmonic oscillators \cite{varios}, but 
it is important to realize that our
problem presents a number of distinctive features that are not found in previous 
treatments, namely

(i) Most of such previous studies have been focused on the coordinates or the momentum
of the oscillator, rather
than on the energy, eg treatments of the time correlation function of the energy are rather
uncommon.

(ii) It is evident that we are dealing with a very special heat bath, since in addition to
being formed by Fermi particles 
 and Bose quasiparticles, such bosons are characterized by a complex dispersion relationship
which gives rise to different species (phonons, rotons R$^+$ and R$^-$).

(iii) The drag force on vortices that has been experimentally detected arise from scattering,
thus we leave aside from our study the phonon radiation damping \cite{arovas,jltp}. 
We note that
such a scattering interaction Hamiltonian is also unusual since it must be nonlinear in the
heat bath operators. In fact, most of the Brownian motion models assume that the heat bath
couples {\em linearly} to the harmonic oscillator, but in our case it is easy to realize that
the scattering events must involve products of creation and annihilation operators of the particles
that collide with the vortex.

(iv) Our recent study \cite{jltp} has shown that the drag force could be affected by appreciable
memory effects, so it will be important to extend our treatment
beyond the usual Markovian approximation to
explore such possible effects.

A suitable formalism to handle the above items can be found in reference \cite{jpa}, where a 
non-Markovian calculation of the energetic susceptibility of a harmonic oscillator, weakly 
coupled to boson and fermion environments, was carried out. So, we shall base our treatment on
the above formalism.
 
This paper is organized as follows, in the following section we describe our quantum
model for the dissipative vortex dynamics which leads to an analysis of vortex oscillations
and memory effects. Next in section 3 we
summarize
the main results of linear response theory applied to the vortex energy. In section 4, based
on the previous calculation of the harmonic oscillator susceptibility, we 
analize memory corrections to the Markov approximation. In section 5 we calculate the response
and time correlation functions, and study the dependence of the relaxation frequency on 
temperature and impurity concentration. Finally, in section 6 we gather the summary and main
conclusions of our study.

\section{Quantum model for vortex dynamics}
We start from a vortex Hamiltonian given by,
\begin{equation}
H_v(z)=\frac{m_v}{2}({\bf v}^2+\Omega\,\omega_-\,{\bf r}^2)\label{p9},
\end{equation}
where
\begin{equation}
{\bf v}={\bf p}/m_v+\frac{\Omega}{2}\,\hat{z}\times {\bf r}
\label{p10}
\end{equation}
corresponds to the vortex velocity ${\bf \dot{r}}$ and ${\bf p}$ denotes the vortex canonical
momentum. The second term in \eref{p10} corresponds to that of the vector 
potential (central gauge) in the electromagnetic analogy, and the $z$ dependence in \eref{p9}
which arises from ${\bf r}(z)$ and ${\bf p}(z)$, corresponds to the rotation in the $x$-$y$ 
plane parametrized by $z$, which results from 
following the helix path. Note that both the canonical momentum ${\bf p}$ and the Hamiltonian
\eref{p9} are given per unit length of the $z$ axis.
Then, it is easy to verify that the Hamilton 
equations
lead from \eref{p9} to the equation of motion \eref{p5}.

The two-dimensional coordinate ${\bf r}$  of the vortex core can be written as the sum of the
center coordinate ${\bf r}_0$ of the cyclotron circle plus the relative coordinate ${\bf r}'$
from such a center. Then, the quantization of such variables straightforwardly arises from
the electromagnetic analogy:
\begin{eqnarray}
{\bf r}_0 & = & \frac{1}{\sqrt{4\pi\rho_sL}}[(\beta^\dagger+\beta)\hat{x}+i(\beta^\dagger-\beta)
\hat{y}]\label{p11}\\
{\bf r}' & = & \frac{1}{\sqrt{4\pi\rho_sL}}[(a^\dagger+a)\hat{x}+i(a-
a^\dagger)\hat{y}],\label{p12}
\end{eqnarray}
where $a^\dagger$ ($\beta^\dagger$) denotes a creation operator of right (left) circular
quanta \cite{cohen} and $L$ denotes 
the vortex line length. The $z$ dependence of ${\bf r}$ arises from the replacements
$\beta^\dagger\rightarrow\exp(-ikz)\,\beta^\dagger_k$,
$a^\dagger\rightarrow\exp(ikz)\,a^\dagger_k$ in \eref{p11} and \eref{p12},
and correspondingly for the 
annihilation operators. Actually, Fetter's theory identifies ${\bf r}_0(z)$ as the whole
displacement from the $z$ axis (see \cite{fet2}, equation (14)). Analogous quantization for the
canonical momentum ${\bf p}$ leads through \eref{p10} and \eref{p9} to a vortex Hamiltonian
\begin{equation}
\fl \int_0^L dz\, H_v(z)=\hbar(\Omega+\omega_-)(a^\dagger_ka_k+\case{1}{2})
+\hbar\omega_-(\beta^\dagger_k\beta_k+\case{1}{2})+
\hbar\omega_-(a^\dagger_k\beta^\dagger_k+a_k\beta_k),\label{p13}
\end{equation}
where it is worthwhile noticing that ${\bf v}^2$ and ${\bf r}^2$ in \eref{p9} turn out to be
independent of $z$, as expected. The above Hamiltonian can be written to first order in 
$\omega_-/\Omega$ as,
\begin{equation}
\hbar\Omega(a^\dagger_ka_k+\case{1}{2})+
\hbar\omega_-(\beta^\dagger_k\beta_k+\case{1}{2})\label{p14}
\end{equation}
where both polarizations (cyclotron and Kelvin modes) become decoupled. To prove this 
approximation, we first note that the set of eigenfunctions of \eref{p14} are represented by 
wave functions corresponding to well-defined values of both numbers of circular quanta, right
and left \cite{cohen}. On the other hand, the Schr\"odinger equation for the Hamiltonian
\eref{p13}, can be easily solved by noting that the term proportional to ${\bf r}^2$
in the expression \eref{p9}, can be added to the corresponding term arising from ${\bf v}^2$,
yielding a Schr\"odinger equation formally equivalent to the one with $\omega_-=0$,
whose solution is well-known. Thus, we find that to the first order in $\omega_-/\Omega$ we
obtain the same spectrum of eigenvalues and eigenfunctions as from \eref{p14}, except for
a slight correction in the radial coordinate of the wave functions, which has to be 
multiplied by the factor  $1+\omega_-/\Omega$.

The Hamiltonian \eref{p14} corresponds to helical oscillations of fixed wavelength $\lambda
=2\pi/|k|$. In the final step of our quantization procedure we shall assume that the system
obeys periodic boundary conditions over a length $L$ along the $z$ axis, so $k$ will be 
restricted to values $2\pi s/L$, where $s$ is a positive or negative integer. Thus, the complete
vortex Hamiltonian is obtained by summing up the expression
\eref{p14} over all these values of $k$:
\begin{equation}
H_v=\sum_k\hbar\Omega(k)(a^\dagger_ka_k+\case{1}{2})+
\hbar\omega_-(k)(\beta^\dagger_k\beta_k+\case{1}{2}).\label{p15}
\end{equation}
The above Hamiltonian differs from the one of Fetter's theory 
 by the presence of the cyclotron modes of frequency $\Omega(k)\simeq\Omega$
(cf \cite{fet2}, equation (11)).

The heat bath Hamiltonian is given by,
\begin{equation}
H_B=\sum_{ {\bf k}}\,\hbar\omega_k \,\,
b_{{\bf k}}^\dagger \,  b_{ {\bf k}}+
\sum_{ {\bf q},\sigma}\,\epsilon_q \,\,
c_{{\bf q},\sigma}^\dagger \,  c_{ {\bf q},\sigma}
\label{hb},
\end{equation}
where $b_{{\bf k}}^\dagger$ denotes a creation operator of $^4$He quasiparticle excitations 
of momentum $\hbar{\bf k}$ and frequency $\omega_k$, and $c_{{\bf q},\sigma}^\dagger$
denotes a creation operator of solvated 
$^3$He atoms of momentum $\hbar{\bf q}$, energy $\epsilon_q$ and
spin 1/2 projection $\sigma$.
Note that we disregard any interaction between the heat bath particles themselves, since we
shall work at low enough temperature and impurity concentration, so that such particles remain
dilute allowing their treatment	 as a noninteracting gas.

To model the scattering interaction Hamiltonian, we will consider a generic momentum-conserving
form:
\begin{equation}
\label{p17}
\int_0^L dz
\sum_{ {\bf k} , {\bf q},\sigma }  \, \, [\Lambda_{{\bf k}  {\bf q}}^{(k)} 
\,b_{{\bf k}}^\dagger \,  b_{ {\bf q}} +
\Gamma_{{\bf k}  {\bf q}}^{(k)}\, c_{{\bf k},\sigma}^\dagger \,  c_{ {\bf q},\sigma}]
e^{-i({\bf k}-{\bf q})\cdot{\bf r}},
\end{equation}
where $\Lambda_{{\bf k}  {\bf q}}^{(k)}$ and $\Gamma_{{\bf k}  {\bf q}}^{(k)}$ denote 
scattering amplitudes depending on the momentum of the heat bath scatterers and the wave vector
$k\hat{z}$ of the vortex wave. Recalling that the vortex coordinate can be written as 
${\bf r}={\bf r}_0(z)+{\bf r}'(z)+z\hat{z}$ and taking into account that ${\bf r}_0(z)$ and
${\bf r}'(z)$ commute, the exponential factor in \eref{p17} can be factorized as
$e^{-i(k_z-q_z)z} e^{-i({\bf k}-{\bf q})\cdot{\bf r}'(z)}
e^{-i({\bf k}-{\bf q})\cdot{\bf r}_0(z)}$. Since the amplitude of the vortex wave was assumed
to be very small, it is tempting to expand the last two exponentials retaining only first order
terms in ${\bf r}'(z)$ and ${\bf r}_0(z)$. This procedure was analyzed by Fetter \cite{fet2}
for ${\bf r}_0(z)$, finding that it leads to divergences at long wavelengths. The physical 
reason for this result can be understood by recalling that ${\bf r}_0(z)$ is linear in the 
creation and destruction operators, $\beta_k^\dagger$ and $\beta_k$. Consequently, an expansion
in powers of ${\bf r}_0(z)$ is bound to fail whenever the energy per quantum $\hbar\omega_-$
becomes very small at long wavelengths, as the transitions should involve many of these ``soft"
quanta \cite{fet2}. Notice that this argument does not apply to ${\bf r}'(z)$ since the operators
$a_k^\dagger$ and $a_k$ correspond to the high-frequency cyclotron quanta.
This means that the treatment of Kelvin modes represented by ${\bf r}_0(z)$ turns out to be
considerably more complicated than that of the cyclotron modes represented by ${\bf r}'(z)$.
In fact, only the phonon drag force arising from ${\bf r}_0(z)$ could be analyzed by Fetter
\cite{fet2}, but we shall see that all sources of friction acting on ${\bf r}'(z)$ can be 
studied. To this aim, let us set ${\bf r}_0(z)=0$ in \eref{p17} while retaining only the 
first order term in ${\bf r}'(z)$. Then, using the second-quantized expression for ${\bf r}'(z)$
(cf \eref{p12}), performing the integral in $z$ and recalling the above-mentioned periodic
boundary conditions, the interaction \eref{p17} reads
\begin{eqnarray}
\fl L\sum_{ {\bf k} , {\bf q},\sigma }  \, \, [\Lambda_{{\bf k}  {\bf q}}^{(k)} 
\,b_{{\bf k}}^\dagger \,  b_{ {\bf q}} +
\Gamma_{{\bf k}  {\bf q}}^{(k)}\, c_{{\bf k},\sigma}^\dagger \,  c_{ {\bf q},\sigma}]
\{\delta_{k_z,q_z} + \frac{1}{\sqrt{4\pi\rho_sL}}[i(q_x-k_x)(\delta_{k,k_z-q_z}
a_k^\dagger + \delta_{k,q_z-k_z}a_k) \nonumber\\
+(q_y-k_y)(\delta_{k,k_z-q_z}
a_k^\dagger - \delta_{k,q_z-k_z}a_k)]\}\label{p18},
\end{eqnarray}
where the Kronecker-delta factors represent $z$-momentum conservation and the scattering 
amplitudes were assumed to be independent of the sign of $k$, ie the interaction should be the
same for both possible 
directions of a helical deformation. Note that the first term between braces 
in \eref{p18} does not contribute to the interaction, so it should be added to the heat bath
Hamiltonian \eref{hb}. Finally, summing up the expression \eref{p18} over $k$ we obtain the 
interaction Hamiltonian:
\begin{eqnarray}
H_{int}=\sqrt{\frac{L}{4\pi\rho_s}}\,
\sum_{ {\bf k} , {\bf q},\sigma }  \, \, [\Lambda_{{\bf k}  {\bf q}}^{(k_z-q_z)} 
\,b_{{\bf k}}^\dagger \,  b_{ {\bf q}} +
\Gamma_{{\bf k}  {\bf q}}^{(k_z-q_z)}\, c_{{\bf k},\sigma}^\dagger \,  c_{ {\bf q},\sigma}]
\nonumber\\
\times[i(q_x-k_x)(a_{k_z-q_z}^\dagger + a_{q_z-k_z}) 
+(q_y-k_y)(a_{k_z-q_z}^\dagger - a_{q_z-k_z})].\label{p19}
\end{eqnarray}
From \eref{p15} and \eref{p19}, we may realize that each cyclotron mode of unperturbed Hamiltonian
\begin{equation}
H_k=
\hbar\Omega(k)(a^\dagger_ka_k+\case{1}{2}),\label{p20}
\end{equation}
will evolve independently, interacting with the heat bath through the following terms of
\eref{p19}:
\begin{eqnarray}
\fl H_{int}^{(k)}=\sqrt{\frac{L}{4\pi\rho_s}}\,
\left\{\sum_{ {\bf k} , {\bf q},\sigma }^{(+)} [\Lambda_{{\bf k}  {\bf q}}^{(k)} 
\,b_{{\bf k}}^\dagger \,  b_{ {\bf q}} +
\Gamma_{{\bf k}  {\bf q}}^{(k)}\, c_{{\bf k},\sigma}^\dagger \,  c_{ {\bf q},\sigma}]\;
[(q_y-k_y)+i(q_x-k_x)]a_k^\dagger\right.\nonumber\\
 + \left.\sum_{ {\bf k} , {\bf q},\sigma }^{(-)}[\Lambda_{{\bf k}  {\bf q}}^{(k)} 
\,b_{{\bf k}}^\dagger \,  b_{ {\bf q}} +
\Gamma_{{\bf k}  {\bf q}}^{(k)}\, c_{{\bf k},\sigma}^\dagger \,  c_{ {\bf q},\sigma}]\;
[(k_y-q_y)+i(q_x-k_x)]a_k\right\},\label{p21}
\end{eqnarray}
where the $(\pm)$ sign above each summation symbol indicates that only the
 terms with $k_z-q_z=\pm k$
must be considered. We recall, however, that for long wavelengths we have 
$k\ll a^{-1}\sim 1$ \AA$^{-1}$, so it will be valid to neglect 
 $k$ in such $z$-momentum conservation relationships, except at extremely low 
temperatures ($\hbar c_s k_z\sim k_BT$). Thus, $H_{int}^{(k)}$ becomes,
\begin{eqnarray}
\fl H_{int}^{(k)}=\sqrt{\frac{L}{4\pi\rho_s}}\,
\sum_{ {\bf k} , {\bf q},\sigma }\delta_{k_z,q_z} [\Lambda_{{\bf k}  {\bf q}}^{(k)} 
\,b_{{\bf k}}^\dagger \,  b_{ {\bf q}} +
\Gamma_{{\bf k}  {\bf q}}^{(k)}\, c_{{\bf k},\sigma}^\dagger \,  c_{ {\bf q},\sigma}]\;
\{[(q_y-k_y)+i(q_x-k_x)]a_k^\dagger\nonumber\\
 +[(k_y-q_y)+i(q_x-k_x)]a_k\}\label{p22}
\end{eqnarray}
and the time evolution of $a_k^\dagger$ will be ruled by the Hamiltonian given by the
sum of \eref{p20}, \eref{hb} and \eref{p22}. To study the time evolution of the vortex coordinate
${\bf r}'=(x',y')$, it will be convenient to use a complex form $R'=x'+iy'$ since
\begin{equation}
R'(z)=\frac{1}{\sqrt{\pi\rho_sL}}\sum_k e^{ikz}a_k^\dagger\label{p23}
\end{equation}
is simply written as a linear combination of $a_k^\dagger(t)$. In reference \cite{jltp}
we studied the cyclotron dynamics of a rigid rectilinear vortex, this being equivalent to 
considering a single term with $k\rightarrow 0$ in \eref{p23}. We derived, within a weak-coupling
approximation, a non-Markovian equation of motion for the mean value of the vortex position
operator, finding that cyclotron frequency values within the range $0.01-0.03$ ps$^{-1}$ 
lead to
a very good agreement with the experimental determinations of the longitudinal friction
coefficient $D$ (equation \eref{p2}), versus temperature and $^3$He concentration.
We showed that memory effects could represent up to $\sim 10\%$ of the $D$ value as the number
of heat bath scatterers is increased, that is, such effects are found to be increasing with
temperature and impurity concentration.
The scattering amplitudes leading to such results reads as \cite{jltp,ca1},
\begin{eqnarray}
\Lambda_{{\bf k}  {\bf q}}^{(0)}=\frac{2\pi\hbar^2}{m_4Vc_s}\sqrt{\frac{19}{140}|\omega'_k||\omega'_q|}
\label{p24}\\
\Gamma_{{\bf k}  {\bf q}}^{(0)}=\frac{3\hbar^2}{m^*V}\sqrt{\frac{\pi}{32}\sigma_0}(kq)^
{\case{1}{4}}\label{p25},
\end{eqnarray}
where $V$ denotes the volume of the system, $\omega'_k$ denotes
the quasiparticle group velocity and
$\sigma_0$=18.54 \AA, corresponds to
an effective cross section for vortex-$^3$He scattering. Note that these
amplitudes are in fact negligible with respect to the heat bath single-particle levels, 
$L\Lambda_{{\bf k}  {\bf k}}^{(0)}\ll\hbar\omega_k$ and 
$L\Gamma_{{\bf k}  {\bf k}}^{(0)}\ll\epsilon_k$, for experimental sizes \cite{pack} and 
not extremely low temperatures.

The thermal excitation of vortex waves can be made consistent with the above experimental data,
if we assume that the scattering amplitudes $\Lambda_{{\bf k}  {\bf q}}^{(k)}$ and
$\Gamma_{{\bf k}  {\bf q}}^{(k)}$ in \eref{p22} are well approximated by the $k=0$ values,
\eref{p24} and \eref{p25}, respectively. Note that this approximation is similar to the previous
one, $k_z-q_z=\pm k\rightarrow 0$ (below equation \eref{p21}) and also to $\Omega(k)\simeq
\Omega$ in \eref{p20}. Thus, each $a^\dagger_k(t)$ in \eref{p23} will present the same
dissipative evolution as $a^\dagger_0(t)$ ie, the same friction coefficient $D$ should be 
ascribed to all long-wavelength cyclotron modes. This generalizes the previous result 
\cite{fet2,son}, 
that the phonon friction coefficient associated to low-frequency Kelvin modes turns out
to be essentially the same as that of strictly rectilinear vortices.

\section{The vortex energy in linear response theory}
According to the standard framework of linear response theory \cite{datta}, we will assume that,
having the vortex reached thermal equilibrium with the heat bath before $t=0$, a weak perturbing
time dependent scalar field $\lambda(t)$
 is coupled to the vortex Hamiltonian from $t=0$ onward. Then, the
Hamiltonian of the whole system can be written,
\begin{equation}
H(t)=H_v+H_B+H_{int}-\lambda(t)H_v,\label{4}
\end{equation}
where $H_B$ is given by \eref{hb}, the vortex Hamiltonian is given by,
\begin{equation}
H_v=
\hbar\Omega\sum_k\,
(a_k^\dagger \,  a_k+\case{1}{2})
\label{p27}
\end{equation}
and the interaction Hamiltonian is given by,
\begin{eqnarray}
\fl H_{int}=\sqrt{\frac{L}{4\pi\rho_s}}\,
\sum_{k, {\bf k} , {\bf q}, \sigma }\delta_{k_z,q_z} [\Lambda_{{\bf k}  {\bf q}}^{(0)} 
\,b_{{\bf k}}^\dagger \,  b_{ {\bf q}} +
\Gamma_{{\bf k}  {\bf q}}^{(0)}\, c_{{\bf k},\sigma}^\dagger \,  c_{ {\bf q},\sigma}]\;
\{[(q_y-k_y)+i(q_x-k_x)]a_k^\dagger\nonumber\\
 +[(k_y-q_y)+i(q_x-k_x)]a_k\}.\label{p28}
\end{eqnarray}
Then,  the mean value of the vortex
energy can be written to the first order in $\lambda(t)$ as,
\begin{equation}
\langle H_v(t)\rangle=\langle H_v\rangle_{eq} + \int_0^t d\tau\lambda(t-\tau)\alpha(\tau),
\label{7}
\end{equation}
where $\langle H_v\rangle_{eq}=N\hbar\Omega\{[\exp(\hbar\Omega/k_BT)-1]^{-1}+\frac12\}$
corresponds to
the canonical equilibrium value ($N$ = total number of long-wavelength cyclotron modes),
 and the function
 $\alpha(\tau)$ embodies the vortex response to the 
applied field. In particular, for a Dirac delta impulse $\lambda(t)=\tau_0\delta(t-t_0)$,
the above equation yields,
\begin{equation}
\frac{\langle H_v(t)\rangle-\langle H_v\rangle_{eq}}{\tau_0}=\alpha(t-t_0)
\label{8}
\end{equation}
that is, $\alpha(\tau)$ represents the energy displacement from the equilibrium value, per
unit strength of a pulse acting at $\tau=0$.
Now, if we assume a constant field $\lambda(t)=\lambda_0$, the so-called
 {\it response function} \cite{datta} is given by,
\begin{equation}
\Psi(t)\equiv \lim_{\lambda_0\rightarrow 0}[
\langle H_v(t)\rangle-\langle H_v\rangle_{eq}]/\lambda_0=
\int_0^t d\tau\,\alpha(\tau).
\label{9}
\end{equation}
Finally, if the field is oscillatory $\lambda(t)=\lambda_0\cos(\omega t)$ ($t\geq 0$), the {\it
nontransient} regime \cite{datta} can be described by setting $t=\infty$ in the upper limit
of the integral in \eref{7},
\begin{eqnarray}
\langle H_v(t)\rangle_{NT}-\langle H_v\rangle_{eq} & = & \lambda_0
 \int_0^\infty d\tau\cos[\omega(t-\tau)]\,\alpha(\tau)\nonumber\\
& = & \lambda_0\, {\rm Re}[\tilde\alpha(\omega)\exp(-i\omega t)]
\label{10}
\end{eqnarray}
where $\tilde\alpha(\omega)$ may be defined as a complex {\it generalized susceptibility},
which is given by the Fourier-Laplace transform of the pulse response $\alpha(\tau)$,
\begin{equation}
\tilde\alpha(\omega)=\int_0^\infty \exp(i\omega\tau)\,\alpha(\tau)\,d\tau.
\label{11}
\end{equation}
Then, according to \eref{11} and \eref{9}, the static susceptibility $\tilde\alpha(0)$ is given by,
\begin{equation}
\tilde\alpha(\omega\rightarrow 0)=\Psi(t\rightarrow\infty)=\frac{N(\hbar\Omega)^2}{4k_BT}
\left[\sinh\left(\frac{\hbar\Omega}{2k_BT}\right)\right]^{-2},
\label{12}
\end{equation}
where the right-hand side arises from taking into account that $\langle H_v(t\rightarrow
\infty)\rangle$ in \eref{9} corresponds to the canonical distribution of a vortex with a 
Hamiltonian $(1-\lambda_0)H_v$ (or equivalently, a vortex with the Hamiltonian $H_v$
at the effective temperature $T/(1-\lambda_0)$).

As a final remark we note that
 from the susceptibility $\tilde\alpha(\omega)$, one can readily get the equilibrium
time correlation function,
\begin{equation}
C(t)=\frac12\langle H_v(t)H_v(0)+H_v(0)H_v(t)\rangle_{eq}-\langle H_v\rangle^2_{eq}
\label{13}
\end{equation}
via the fluctuation-dissipation theorem \cite{callen}:
\begin{equation}
\bar C(\omega)=\hbar\coth\left(\frac{\hbar\Omega}{2k_BT}\right){\rm Im}[\tilde\alpha(\omega)],
\label{14}
\end{equation}
where $\bar C(\omega)=\int_{-\infty}^\infty dt \exp(i\omega t)C(t)$ denotes the Fourier
transform of $C(t)$.

\section{Analytic continuation of the generalized susceptibility: study of memory effects}
Being a Laplace transform, the generalized susceptibility \eref{11} can be regarded as a 
function of a complex variable $z$, $\tilde\alpha(z)$, which must be analytic in the upper half-plane,
Im $z>0$. The important information, however, lies in the lower half-plane, where the spectrum
of singularities of its analytic continuation yields the set of characteristic frequencies in
the time evolution of the pulse response $\alpha(\tau)$ \cite{jpa}.
Our calculation of $\tilde\alpha(z)$ is completely analogous to the one leading to the harmonic
oscillator susceptibility in reference \cite{jpa}. Thus, we refer the reader to that paper for
the technical details, and only quote here the final result (cf \cite{jpa}, equation (2.25)):
\begin{equation}
\tilde\alpha(z)=\frac{N\,\hbar^2\Omega^2[q(z)-q(0)]/z}{[1-\exp(-\hbar\Omega/k_BT)][z+iv(z)]},
\label{15}
\end{equation}
where $q(z)$ and $v(z)$ are Cauchy integrals,
\begin{eqnarray}
q(z) & = & \frac{1}{2\pi i}\int_{-\infty}^\infty \frac{d\omega}{\omega-z}F(\omega)\label{16}\\
v(z) & = & \frac{1}{2\pi i}\int_{-\infty}^\infty \frac{d\omega}{\omega-z}\nu(\omega)\label{17},
\end{eqnarray}
with kernels (cf \cite{jpa}, equations (2.31) and (2.32)),
\begin{eqnarray}
F(\omega) & = & \frac{R(\omega)n(\Omega+\omega)}{n(\omega)}-
\frac{R(-\omega)n(\Omega-\omega)}{n(-\omega)}\label{18}\\
\nu(\omega) & = & \frac{\hbar}{i}[R(\omega)+R(-\omega)]\label{19},
\end{eqnarray}
being,
\begin{equation}
n(\omega)=[\exp(\hbar\omega/k_BT)-1]^{-1}
\label{20}
\end{equation}
and
\begin{equation}
R(\omega)=\frac{2i}{m_v\hbar\Omega}(\Omega+\omega)D(\Omega+\omega).
\label{21}
\end{equation}
The function $D$ in the above equation has been studied in previous works \cite{ca1,jltp},
since $D(\Omega)$ corresponds to the longitudinal
friction coefficient in the Markovian approximation. It reads \cite{jltp},
\begin{eqnarray}
D(\Omega) & = &
\frac{L\pi}{2\hbar\Omega}\sum_{ {\bf k}, {\bf q}}  \, \, \delta_{k_zq_z}
({\bf k}-{\bf q})^2[|\Lambda_{{\bf k}  {\bf q}}^{(0)}|^2(n_q-n_k)
\delta(\omega_k-\omega_q-\Omega)\nonumber\\
&&+2|\Gamma_{{\bf k}  {\bf q}}^{(0)}|^2(f_q-f_k)
\delta(\epsilon_k/\hbar-\epsilon_q/\hbar-\Omega)],
\label{22}
\end{eqnarray}
where $n_k=[\exp(\hbar\omega_k/k_BT)-1]^{-1}$ and $f_k=\{\exp[(\epsilon_k-\mu)
/k_BT]+1\}^{-1}$ respectively denote the thermal equilibrium Bose and Fermi occupation numbers
for the corresponding scatterers. 
The expression \eref{15} for the susceptibility, on the other hand, is fully non-Markovian and
each pole $z_j$ of it yields a term proportional to $\exp(-iz_j\tau)$ in $\alpha(\tau)$.
Then, the Markov approximation consists in neglecting the set of such poles which are located
far enough from the origin, so that they yield rapidly vanishing terms, ie terms which decay
faster than any observational timescale. This is the case for the set of poles arising from
$q(z)$ in \eref{15}, actually poles of $n(\Omega\pm\omega)$ in \eref{18} \cite{jpa}, which
are of the form $\pm\Omega-in2\pi k_BT/\hbar$ ($n=$1,2,...). Such poles give rise to 
exponentially decaying terms in the expression of $\alpha(\tau)$, which have lifetimes shorter
than $\hbar/k_BT$. This thermal timescale turns out to be much smaller than the one arising
from the friction coefficient, ie $\hbar/k_BT\ll m_v/D(\Omega)=[\rho_sh/D(\Omega)]\Omega^{-1}$
for $T<1.5$ K and $\Omega$'s of order
0.01 ps$^{-1}$. So, the above set of thermal poles can be safely ignored. It is clear then,
that we should look for poles of order $\Omega[D(\Omega)/\rho_sh]$ and, in the following,
we shall see that they arise from the equation $z+iv(z)=0$. In fact, looking for a solution
close to the origin, one may begin with the ansatz $z_0=-iv(0)$ and next proceed iteratively,
ie $z_1=-iv(z_0)$ and so on. This leads to a very rapid convergence to a solution $z_s$ that
is better worked out in terms of the Taylor expansion of $v(z)$ around the origin:
\begin{equation}
z_s=-iv(0)[1-iv'(0)-v''(0)v(0)/2],
\label{23}
\end{equation}
where the second and third term inside the square brackets represent first and second order
corrections to the zeroth order solution, respectively. The Cauchy integral \eref{17} and
its derivatives in \eref{23}, can be written as \cite{jpa},
\begin{equation}
v(0)=\nu(0)/2=2\Omega D(\Omega)/\rho_sh
\label{24}
\end{equation}
\begin{eqnarray}
v'(0) & = & \frac{1}{2\pi i}\int_{-\infty}^\infty \frac{d\omega}{\omega^2}
[\nu(\omega)-\nu(0)]\nonumber\\
& = & \frac{2}{i\pi\rho_sh}\int_0^\infty \frac{d\omega}{\omega^2}
[(\Omega+\omega)D(\Omega+\omega)+(\Omega-\omega)D(\Omega-\omega)\nonumber\\
& - & 2\Omega D(\Omega)]\label{25}
\end{eqnarray}
\begin{equation}
v''(0)=\nu''(0)/2=\frac{2}{\rho_sh}[2D'(\Omega)+\Omega D''(\Omega)].
\label{26}
\end{equation}
We see from \eref{24} that the zeroth-order solution $z_0=-iv(0)$ has in fact the expected dependence
and, moreover, the factor 2 in the expression of $v(0)$ is easily interpreted if we  recall
that a damping in the velocity like $\exp(-\Omega t D(\Omega)/\rho_sh)$ should give rise to
a twice faster energy damping. The first and second order corrections in \eref{23} are easily
evaluated from \eref{25} and \eref{26}, and they are always negligible, eg for ordinary helium
at $T=0.67$ K ($\Omega=0.01$ ps$^{-1}$) we have $-iv'(0)=1.52\times 10^{-5}$ and 
$-v''(0)v(0)/2=1.32\times 10^{-9}$. In conclusion, we have found that the Markovian pole
$-i2\Omega D(\Omega)/\rho_sh$ is unaffected by memory corrections. This is to be contrasted
with the appreciable memory corrections to the friction coefficient seen in section 2.
 Therefore, the Markovian approximation for the generalized
susceptibility \eref{15} reads,
\begin{equation}
\tilde\alpha_M(z)=\frac{N\,\hbar^2\Omega^2 q'(0)}{[1-\exp(-\hbar\Omega/k_BT)][z+iv(0)]}
\label{27}
\end{equation}
which should be valid for $z$ inside a circle with a radius $r_0$ fulfilling $v(0)<r_0\ll
k_BT/\hbar$. Notice that we have replaced the expression $[q(z)-q(0)]/z$ in \eref{15} by the
derivative $q'(0)$. This approximation may be readily tested if one considers the first order
term $q''(0)z/2$ at $z=z_0=-iv(0)$. In fact, we have \cite{jpa}
\begin{equation}
q'(0)=F'(0)/2=-n(\Omega)z_0/k_BT
\label{28}
\end{equation}
and
\begin{equation}
q''(0)=\frac{2}{\pi i}\int_0^\infty \frac{d\omega}{\omega^3}
[F(\omega)-F'(0)\omega],
\label{29}
\end{equation}
so, we may find again that the first order correction is totally negligible, eg for ordinary
helium at $T=0.67$ K ($\Omega=0.01$ ps$^{-1}$) it represents $\sim 10^{-6}$ of the zeroth
order $q'(0)$. Then, replacing \eref{28} in \eref{27} we have the final expression,
\begin{equation}
\tilde\alpha_M(z)=\frac{N\,\hbar^2\Omega^2\, i2\Omega D(\Omega)/\rho_sh}
{4k_BT\sinh^2(\hbar\Omega/2k_BT)\,[z+i2\Omega D(\Omega)/\rho_sh]}
\label{30}
\end{equation}
which immediately reproduces the result \eref{12} for the static susceptibility $\tilde
\alpha(0)$.

\section{Response and time correlation functions}
From \eref{30} one easily extracts $\alpha(\tau)$ and the response function \eref{9},
\begin{equation}
\Psi(t)=\frac{N\,(\hbar\Omega)^2}{4k_BT}
\left[\sinh\left(\frac{\hbar\Omega}{2k_BT}\right)\right]^{-2}
[1-\exp(-2\Omega t D(\Omega)/\rho_sh)].
\label{31}
\end{equation}
This kind of response, characterized by a single relaxation time, is well-known in theories
of dielectric and magnetic relaxation and goes under the name of {\it Debye response} \cite
{datta}.

The Fourier transform of the time correlation function arises from \eref{14} and \eref{30}:
\begin{equation}
\bar{C}(\omega)=\frac{N\,(\hbar\Omega)^2}{4k_BT}
\left[\sinh\left(\frac{\hbar\Omega}{2k_BT}\right)\right]^{-2}
\frac{\hbar\coth(w)w\epsilon}{\epsilon^2+w^2},
\label{32}
\end{equation}
where $\epsilon=[\Omega D(\Omega)/\rho_sh]/[k_BT/\hbar]$ and $w=\hbar\omega/2k_BT$.
Being $\epsilon\ll1$, the function of $w$, $\epsilon/(\epsilon^2+w^2)$ in \eref{32} turns out
to be sharply peaked around $w=0$, so we may approximate $\coth(w)w\simeq 1$ and thus get
the antitransform:
\begin{equation}
C(t)=\frac{N\,(\hbar\Omega)^2\exp[-2\Omega |t|D(\Omega)/\rho_sh]}
{4\sinh^2(\hbar\Omega/2k_BT)}.
\label{33}
\end{equation}
Therefore, the time correlation function is ruled by the same relaxation time of the 
response function, as expected. Notice that $C(0)$ in \eref{33} actually corresponds to
$\langle H_v^2\rangle_{eq}-\langle H_v\rangle^2_{eq}$, as can be easily verified by an
elementary calculation of $\langle H_v^2\rangle_{eq}$ in the canonical ensemble.

To conclude, it is interesting to analyze how the relaxation frequency $2\Omega D(\Omega)/
\rho_sh$ depends on temperature and impurity concentration. In figure 2, $D(\Omega)/\rho_sh$
\begin{figure}
\begin{center}
\epsfbox{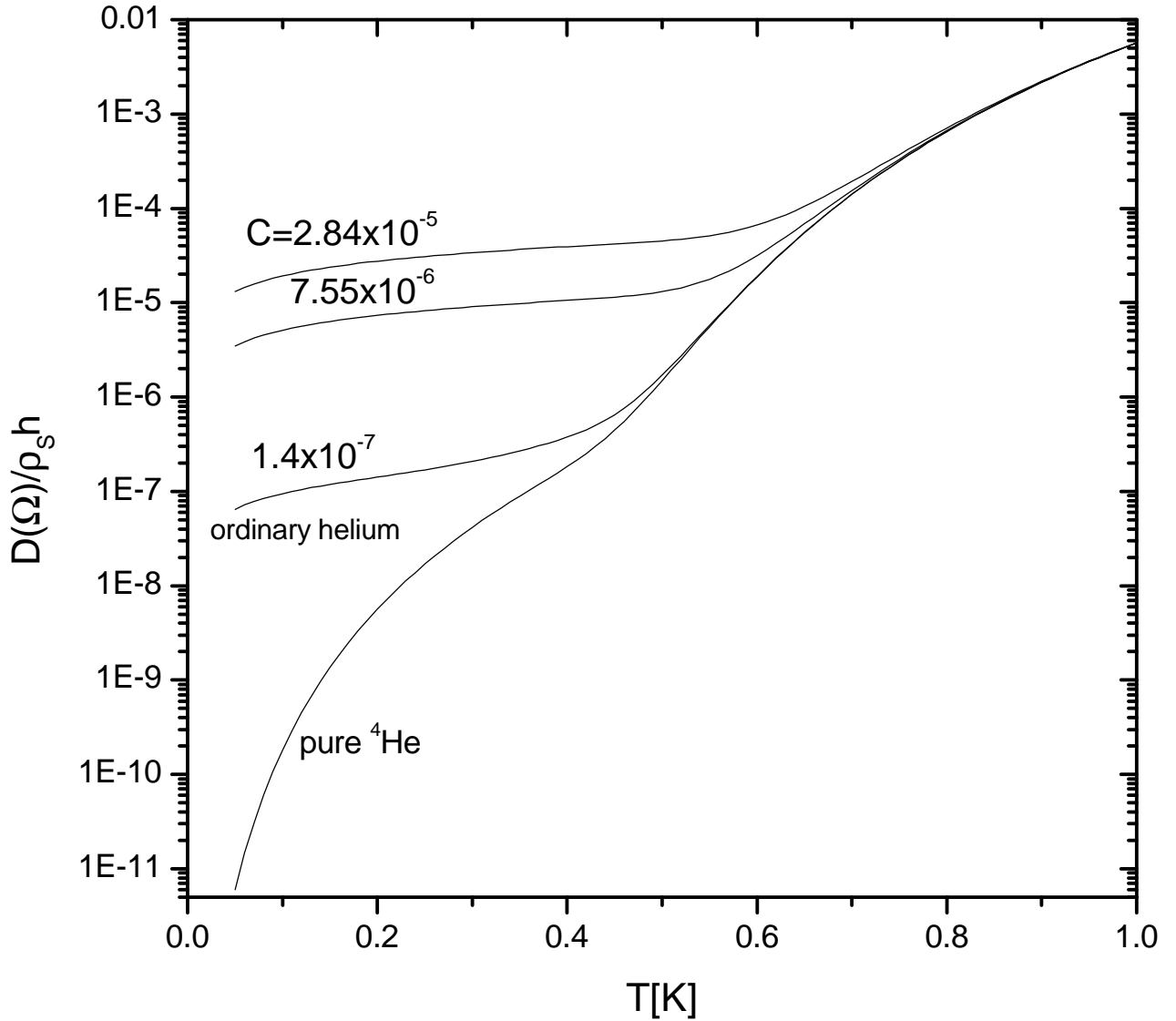}
\end{center}
\caption{\label{fig2}Relative value of the friction coefficient $D(\Omega)$ with respect to
the Magnus force coefficient $\rho_sh$, versus temperature for several $^3$He 
concentrations $C$.}
\end{figure}
is plotted against temperature for several $^3$He concentrations \cite{reif}. 
Such curves actually correspond to $\Omega=0.01$ ps$^{-1}$, but it is important to remark
that the dependence on $\Omega$ turns out to be negligible for $\Omega$'s within $10^{-2}$
ps$^{-1}$
or less \cite{ca1,jltp}. It is also worth noticing that consistently with our weak-coupling 
approximation \cite{ca1,jltp,jpa}, the Markovian
  friction coefficient $D(\Omega)$ always remains small
compared to the coefficient $\rho_sh$ of the Magnus force in \eref{1} (actually $\rho_sh$
has virtually no dependence on temperature for $T<1$ K). The lowest curve in figure 2
corresponds to pure $^4$He and it displays two well separated regimes \cite{ca1},
\[D(\Omega) \sim \left\{\begin{array}{ll}
 T^5, & \mbox{for $T<0.4$ K (phonon domain)}\\
\exp(-\Delta/k_BT), &
\mbox{for $T>0.5$ K (roton domain),}
\end{array}
\right.\]
where $\Delta/k_B$=8.62 K corresponds to the height of the roton minimum in the dispersion
curve of figure 1. Note that the phonon-roton transition clearly manifests itself as an
intermediate region of positive second derivative. The remaining curves in figure 2 
correspond to finite $^3$He concentrations that have been experimentally studied \cite{reif}.
Such concentrations are low enough to allow a Maxwell-Boltzmann approximation for the $^3$He
statistics in \eref{22}. Then, the low-temperature regime of $D(\Omega)$, which is now 
dominated by impurity scattering, turns out to be proportional to $\sqrt{T}$ and $^3$He
concentration \cite{reif,jltp}. We may see from figure 2 that phonon effects are completely
hidden in ordinary helium, since the $^3$He domain extends as far as $T\simeq 0.4$ K.
For higher concentrations such domain reaches higher temperatures hiding also the first
portion of the roton curve.

Finally, it is worthwhile observing that relaxation frequencies with any temperature dependence,
rarely appear in models of quantal Brownian motion of harmonic oscillators, since most of them
assume, in contrast to our scattering model, linear couplings in the heat bath operators
\cite{varios,jpa}.

\section{Summary and conclusions}
A generalization of the quantum theory of vortex waves \cite{fet1,fet2}, has been proposed
to study the damping of cyclotron modes. We have shown that the friction values should be
practically unaffected by such oscillations, a result which was already known in the case of
phonon scattering of low-frequency Kelvin modes \cite{fet2,son}. All sources of dissipation
arising in ordinary helium, viz phonons, rotons and $^3$He atoms were considered, showing
that appreciable memory effects must be taken into account in the evaluation of the friction
coefficient. We have also analyzed memory corrections to the Markov approximation in the case
of the nonequilibrium energetics of cyclotron modes, finding this time that they are negligible.
We have shown that the vortex response
is of the Debye type, ie it is ruled by a single relaxation frequency which governs
the time correlation function as well. 
Such a relaxation frequency is shown to embody
all the complexity of the heat bath, in that very well separated regimes belonging to the 
different species comprising the superfluid helium, are recognized from its
dependence on temperature and impurity concentration.

\ack
This work has been performed under Grant PEI 6370 from CONICET, Argentina.

\end{document}